\documentclass[twocolumn]{revtex4}

\usepackage{amsmath}    
\usepackage{amssymb}
\usepackage{graphicx}   

\begin{document}


\title{The evolution of free wave packets}

\author{Mark Andrews}

\email{Mark.Andrews@anu.edu.au}
\affiliation{Department of Physics, Australian National University, ACT 0200, Australia} 
\date{\today}

\begin{abstract}
We discuss four general features of force-free evolution: (1) The spatial spread of any packet changes with time in a very simple way. (2) Over sufficiently short periods of time (whose duration is related to the spread in momentum of the packet) the probability distribution moves but there is little change in shape. (3) After a sufficiently long period (related to the initial spatial spread) the packet settles into a simple form simply related to the momentum distribution in the packet. In this asymptotic regime, the shape of the probability distribution no longer changes except for its scale, which increases linearly with the time. (4) There is an infinite denumerable set of simple wave packets (the Hermite-Gauss packets) that do not change shape as they evolve. 
\end{abstract}

\maketitle

\section{Introduction}

The behavior of free wave packets as they evolve in time is striking, even alarming, to many students of quantum mechanics. A wave packet usually changes shape and always eventually spreads out without limit. How different this is from the behavior of a free classical particle! Reconciliation of the classical and quantum descriptions is part of the bigger question of the interpretation of quantum mechanics, but the dynamics of free wave packets is still of considerable interest in introductory quantum mechanics. 

Most texts derive the propagator
\begin{equation}
\label{eq:prop}
K(x,x',t)=\sqrt{m/2\pi \imath \hbar t}\,\exp [\imath m (x-x')^2 / 2\hbar t],
\end{equation}
such that the evolution of an arbitrary initial wave packet $\psi (x,0)$ is exactly
\begin{equation}
\label{eq:intProp}
\psi(x,t)=\int _{-\infty}^{\infty}K(x,x',t)\psi (x',0)\,dx'.
\end{equation}
Unfortunately, there are not many wave packets for which this integral can be easily evaluated and the only case commonly treated is the Gaussian packet. Therefore the evolution of free wave packets remains obscure to many students.

Ehrenfest's result that $\langle \hat p \rangle$ is constant and that $\langle \hat x \rangle =\langle \hat x \rangle_{0}+\langle \hat p \rangle t/m$ is well known. There are four other insights into the free evolution of wave packets that are simply derived and greatly improve the general understanding of this evolution: 

\textbf{1}. The spread of any wave packet, as measured by $\Delta _{x} = \langle (\hat x-\langle \hat x \rangle)^{2}\rangle^{1/2}$, varies in time as
\begin{equation}
\label{eq:delx}
 \Delta _{x} =  \sqrt{\Delta_{min}^2+(t-t_{min})^{2} \Delta _{p}^{2} / m^{2}},
\end{equation}
where $\Delta _{min}$ is the minimum spread and $\Delta _{p}=\langle (\hat p-\langle \hat p \rangle)^{2}\rangle^{1/2}$ does not change with time.  If the wave function is real everywhere (which can be only for an instant) then the packet has its minimum spread $\Delta_{min}$ at that instant.

\textbf{2}. Over sufficiently short periods of time, much less than $ m\hbar /\Delta _{p}^{2}$, the wave packet moves (with speed $\langle \hat p \rangle/m$) without much change in the shape of the probability distribution.

\textbf{3}. The asymptotic evolution, for times $|t|\gg m\Delta _{x}^2 /\hbar$, is to a form simply related to the momentum distribution in the packet. This means that for any initial packet for which the momentum distribution (the Fourier transform of the wave function) can be easily evaluated, the evolution of the packet in this asymptotic period is accessible.

Note that these two time regimes can not overlap, because of Heisenberg's uncertainty relation $\Delta _{x}\Delta _{p}\geqslant \hbar/2$.

\textbf{4}. There is an infinite number of simple wave packets that do not change shape (apart from the inevitable spreading) as they evolve. These wave functions have the form of a Gaussian multiplied by a polynomial in $x$. 

In a recent note\cite{Mita}, it is claimed that ``a free particle wave packet of any shape becomes approximately Gaussian after a period of time''. This claim is false. An obvious counter-example is provided by any odd wave packet: it will remain odd as it evolves freely and therefore can never approximate a Gaussian. Furthermore, the claim is not consistent with Properties 3 and 4 above. This note is discussed further in the appendix. Here we will establish the four properties and give some examples.

We use the term \textit{wave packet} to indicate that the wave function is sufficiently confined in space for $\langle \hat{x}^{2}\rangle$ to exist.

\section{The evolution over short periods}

Starting with the well-known formula\cite{Rob} for the evolution in terms of an initial momentum distribution $\phi(p)$,
\begin{equation}
\label{eq:psip}
\psi(x,t)=\frac{1}{\sqrt{2\pi \hbar}}\int_{-\infty}^{\infty} \exp [\frac{\imath}{\hbar} (p x-\frac{1}{2m}p^{2}t)]\phi(p)\,dp,
\end{equation}
expand\cite{Ohan} $p^{2}$ about the mean value $\bar{p}=\langle \hat p \rangle$. Thus $p^{2}=\bar{p}^{2}+2\bar{p}p +(p-\bar{p})^{2}$. We expect $\phi(p)$ to be small if $(p-\bar{p})^2\gg\Delta_{p}^2$ and therefore, if $\Delta_{p}^{2}t/2m\hbar \ll 1$, the term in $(p-\bar{p})^{2}$ will make a negligible change to the phase of the exponential. In other words, the integrand contains a factor $\exp [\imath (p-\bar{p})^{2}t/2m\hbar]$ that will be very close to unity throughout the region where $\phi(p)$ has any significant magnitude. Hence 
\begin{eqnarray} \label{eq:psipa}
\psi(x,t) & \approx &\!\frac{1}{\sqrt{2\pi \hbar}} \exp ( \frac{-\imath\bar{p}^{2}t}{2m\hbar})\!\!\int_{-\infty}^{\infty}\!\!\!\!\exp [\frac{\imath}{\hbar} p(x-\frac{\bar{p}t}{m})]\phi(p)\,dp,\nonumber \\
& = & \exp ( \frac{-\imath\bar{p}^{2}t}{2m\hbar})\, \psi(x-\frac{\bar{p}t}{m},0).
\label{eq:early}
\end{eqnarray}
The wave function translates, with speed $\bar{p}/m$, and no change in shape, just a change in phase. A more rigorous approach to the error in this approximation is given in Section VII.

Here we have taken the initial time to be $t=0$, but this analysis can be applied at any time. Since $\Delta_{p}$ does not change, the time scale $m\hbar/2\Delta_{p}^{2}$ for changes in shape applies throughout the evolution.

If the wave function has any discontinuity then $\Delta_{p}=\infty$ and there will be very rapid changes. An example of this behavior will be considered in Section VI.

\section{The evolution for large times}

For the propagator in Eq.~(\ref{eq:prop}), write the term $(x-x')^2$ in the exponent as $x^{2}-\bar{x}^{2}-2(x-\bar{x})x'+ (x'-\bar{x})^{2}$, with $\bar{x}=\langle \hat x \rangle_{t=0}$. We expect $\psi(x',0)$ to be small if $(x'-\bar{x})^2\gg\Delta_{x}^2$ and therefore, for all times (future or past) with $|t|\gg m\Delta_{x}^2 /\hbar$, the term in $(x'-\bar{x})^{2}$ will make a negligible change to the phase of the exponential. Hence 
\begin{eqnarray} \label{eq:large_t}
 \psi(x,t) & \approx & \sqrt{\frac{m}{2\pi \imath \hbar t}}\, \exp \big[\frac{\imath m}{2\hbar t}(x^2-\bar{x}^{2})\big]\times\nonumber\\
 & &\int_{-\infty}^{\infty} \exp \big[-\frac{\imath m}{\hbar t} (x-\bar{x}) x'\big]\psi(x',0)\,dx'\nonumber \\
&  &  \hspace{-1.5cm}= \sqrt{\frac{m}{ \imath t}}\, \exp \big[\frac{\imath m}{2\hbar t}(x^2-\bar{x}^{2})\big]\phi \big(m(x-\bar{x})/t\big), 
\end{eqnarray}
where \begin{equation}
\label{eq:mom}
\phi(p)=\frac{1}{\sqrt{2\pi \hbar}}\int_{-\infty}^{\infty} \exp (-\imath p x'/\hbar)\psi(x',0)\,dx',
\end{equation}
which is the initial momentum wave function of the packet. A more rigorous derivation of the error in this approximation is given in Section VII.  Historically, this asymptotic form was important in the theory of scattering.\cite{scat} 

The asymptotic probability distribution is therefore $(m/t)\,|\phi\big(m(x-\bar{x})/t\big)|^{2}$. This has a simple interpretation: $m(x-\bar{x})/t$ is the momentum required for the particle to be at position $x$ at time $t$, if it left $\bar{x}$ at time $t=0$. So the probability of finding the particle at $x$ is proportional to the probability that it had the right momentum to get there (ignoring the fact that the initial distribution was spread over a distance of order $\Delta_{x}$ about $\bar{x}$).

Introduce the times $t_{p}=m\hbar/2\Delta_{p}^{2}$ and $t_{x}=2m\Delta_{x}^{2}/\hbar$. Then $t_{p}\leqslant t_{x}$ (from the uncertainty relation) and equality implies a Gaussian packet. In general, $t_{p}$ gives the time scale for changes in shape (including the changes in spatial scale), and $t_{x}$ gives the time scale to move to the asymptotic period where there is no further change in shape but $|\psi |$ expands uniformly with time.

\section{The evolution of the spread $\Delta_{x}$}

The fact that every free packet spreads exactly as in Eq.~(\ref{eq:delx}) has appeared in at least one textbook\cite{Merz}, but the derivation is simpler if the concept of the \textit{total time derivative} of an operator is used\cite{And}. The total time derivative, $d_{t}\hat A$ of any operator $\hat A$ is defined to be 
\begin{equation}
\label{dtA}
d_{t}\hat A=\partial_{t}\hat A +(\imath /\hbar)[\hat H,\hat A],
\end{equation}
where $\hat H$ is the Hamiltonian operator. For the free particle this leads to the operator equations of motion $md_{t}\hat x = \hat p$ and $d_{t}\hat p = 0$. From the general property that $d_{t}\langle \hat A\rangle=\langle d_{t}\hat A\rangle$, it follows that $md_{t}\hat X = \hat P$ and $d_{t}\hat P = 0$, where $\hat X=\hat x-\langle \hat x\rangle$ and $\hat P=\hat p-\langle \hat p\rangle$. Furthermore, $md_{t}\hat X^{2} = 2 \hat R$, where $\hat R=(\hat P\hat X +  \hat X\hat P)/2$, and $md_{t}\hat R=\hat P^{2}$. Since $\langle \hat P^{2}\rangle$ is constant, it follows that 
\begin{equation}
\label{eq:mR}
m\langle\hat R\rangle=\langle\hat P^{2}\rangle t +m\langle\hat R\rangle_{t=0}=\langle\hat P^{2}\rangle(t-t_{min}),
\end{equation}
where $t_{min}=-m\langle\hat R\rangle_{t=0}/\langle\hat P^{2}\rangle$. Now we can integrate $md_{t}\langle \hat X^{2}\rangle = 2\langle\hat R\rangle$ to give 
\begin{equation}
\label{eq:X2}
\langle \hat X^{2}\rangle =m^{-2}\langle\hat P^{2}\rangle(t-t_{min})^{2}+\langle \hat X^{2}\rangle_{min},
\end{equation}
which is equivalent to Eq.~(\ref{eq:delx}). Thus, for any wave packet, one can calculate the time of minimum spread from $\langle\hat P^{2}\rangle$ and $\langle\hat R\rangle_{t=0}$. A wave packet has its minimum spread at the time when $\langle\hat R\rangle=0$.

If the wave function $\psi$ is real at some time, the $\imath$ in Schr\~odinger's equation $\partial_{t}\psi=-\imath (\hbar/2m) \partial_{x}^{2}\psi$ shows that $\psi$ will immediately become complex. Also $\langle \hat p \rangle=0$ because $\langle \hat p \rangle$ = $\int \!\psi \,\hat p \psi \,dx$ = $\int (\hat p \psi)^{\ast}\psi \,dx$ = $-\int (\hat p\psi) \psi \,dx$ = $-\langle \hat p \rangle$.  Similarly, $\langle \hat p\hat x \rangle$ = $\int \!\psi \,\hat p \hat x\psi \,dx$ = $\int (\hat p \psi)^{\ast}\hat x\psi \,dx$ = $-\int (\hat p\psi ) \,(\hat x\psi) \,dx$=$-\langle \hat x\hat p \rangle$. That is, $\langle \hat R \rangle=0$ and therefore the wave packet has its minimum spread at that instant. This also shows that a free wave packet can be real only for one instant. In fact, the probability is symmetric in time about that instant, because if $\psi(x,t)$ satisfies Schr\~odinger's equation then so does $\psi^{*}(x,2t_{0}-t)$ and since these two functions are equal at $t=t_{0}$ if $\psi(x,t_{0})$ is real, they must be equal at all times.

Given a wave packet $\xi$ that is at rest (in the sense that $\langle \hat p \rangle_{\xi}=0$ and therefore $\langle \hat x \rangle_{\xi}$ is constant) we can make a moving packet by multiplying the wave function by $\exp(\imath p x/\hbar)$. With $\psi(x,t_{0})= \exp(\imath p x/\hbar)\xi(x,t_{0})$, it follows that $\hat p\, \psi=p\,\psi+\exp(\imath p x/\hbar)\hat p \,\xi$ and therefore $\langle \hat p \rangle_{\psi}=p$. Similarly, $\langle \hat p^{2} \rangle_{\psi}=p^{2}+\langle \hat p^{2} \rangle_{\xi}$ and hence $(\Delta_{p})_{\psi}=(\Delta_{p})_{\xi}$. Also, $\hat p\,(x \psi)=p\,x\psi+\exp(\imath p x/\hbar)\hat p \,(x\xi)$ and therefore $\langle \hat p \hat x+\hat x \hat p\rangle_{\psi}=2p\langle \hat x \rangle_{\xi}+\langle \hat p \hat x+\hat x \hat p\rangle_{\xi}$. Thus,  $\langle \hat R \rangle_{\psi}=\langle \hat R \rangle_{\xi}$ and it follows that the evolution of the spatial spread is the same as for $\xi$. The evolution of $\psi$ can be obtained from that of $\xi$ by applying a Galilean transformation\cite{Merz2}:
\begin{equation}
\label{eq:Galileo}
\psi(x,t)=\exp\big(\frac{\imath}{\hbar}(px-p^{2}\frac{t-t_{0}}{2m})\big)\,\xi(x-p\frac{t-t_{0}}{2m},t).
\end{equation}
It is easy to check that this $\psi$ satisfies  Schr\~odinger's equation if $\xi$ does, and that $\psi(x,t_{0})=\exp(\imath px/\hbar)\,\xi(x,t_{0})$.
  
\section{The Hermite-Gauss wave packets}

A vast array of time-dependent solutions of Schr\~odinger's equation (SE) can be found through the use of \textit{invariant operators}\cite{And}. For any operator $\hat A$, 
\begin{equation}
\label{eq:invOp}
(\hat H-\imath \hbar \partial_{t})\hat A \psi =([\hat H,\hat A]-\imath \hbar \,\partial_{t}\hat A)\psi=-\imath\hbar (d_{t}\hat A) \psi.
\end{equation}
If $d_{t}\hat A=0$ at all times, we say that $\hat A$ is \textit{invariant}. If an invariant operator is applied to any solution of SE, then from Eq.~(\ref{eq:invOp}) the result will also satisfy SE. After being multiplied by a suitable time-dependent factor (if necessary), eigenfunctions of invariant operators will satisfy SE and the eigenvalue will be constant.  [If the eigenfunctions are degenerate, appropriate linear combinations, with time-dependent coefficients, may be required.] 

In the present case of a free particle, a useful invariant operator is 
\begin{equation}
\label{eq:b}
\hat b=m \hat x - \hat p (t-\imath \tau),\,\,\,\,\tau \,\,\text{real}
\end{equation}
The eigenfunction of $\hat b$ with eigenvalue zero (so that $\langle \hat x \rangle = \langle \hat p \rangle = 0$) is easily found using $\hat p = -\imath\hbar\partial_{x}$:
\begin{equation}
\label{eq:chi}
\chi (x,t)=(\frac{m \tau}{\pi \hbar})^{1/4}\frac{1}{ \sqrt{ (t-\imath \tau)}}\exp (\frac{\imath m  x^{2}}{2\hbar (t-\imath \tau)}).
\end{equation}
This is Schr\~odinger's well-known free Gaussian wave packet. The factor $(t-\imath \tau)^{-1/2}$ does not come from the eigenvalue equation $\hat b\, \chi = 0$, but it is clear from the propagator in Eq.~(\ref{eq:prop}) that this factor is required to satisfy SE. Propagators always satisfy SE and with $x'=0$ the propagator in Eq.~(\ref{eq:prop}) has the same form as $\chi$ with $\tau=0$. Furthermore, any solution of SE is still a solution if a constant (such as $-\imath\tau$) is added to $t$; thus $\chi$ satisfies SE.

Applying $\hat b^\dag=m \hat x - \hat p (t+\imath \tau)$ to the eigenfunction $\chi$ will give another solution to SE (because $\hat b^\dag$ is also invariant), and clearly this must have the form of the same Gaussian multiplied by $x$. Applying $\hat b^\dag$ again gives a solution that has the form of the same Gaussian multiplied by a quadratic polynomial in $x$, and so on. We will now show that the sequence of polynomials generated in this way are Hermite polynomials. 

First write $\exp [\imath m x^{2}/2\hbar (t-\imath\tau)]$ in Eq.~(\ref{eq:chi}) as $\exp [\imath m x^{2}/2\hbar (t+\imath\tau)]\,\exp [-m \tau x^{2}/\hbar (t^{2}+\tau^{2})]$ and then, since $\hat b^\dag =m x+\imath\hbar (t+\imath\tau)\partial_{x}$, 
\begin{equation}
\label{eq:applyb}
\hat b^\dag \exp [\frac{\frac{1}{2}\imath m x^{2}}{\hbar (t+\imath\tau)}]\,f =\imath\hbar (t+\imath\tau) \exp [\frac{\frac{1}{2}\imath m x^{2}}{\hbar (t+\imath\tau)}]\,\partial_{x}f 
\end{equation}
for any function $f(x,t)$. Now we can apply this result repeatedly, starting with $\chi$, to give
\begin{equation}
\label{eq:chin}
\chi_{n}(x,t)\propto \frac{(t+\imath\tau)^{n}}{ \sqrt{ (t-\imath \tau)}}\exp [\frac{\frac{1}{2}\imath m x^{2}}{\hbar (t+\imath\tau)}]\,\partial_{x}^{n}\,\exp (-\frac{ x^{2}}{\gamma^{2}}),
\end{equation}
where we have introduced the time-varying length scale $\gamma = [\hbar (t^{2}+\tau^{2})/m\tau]^{1/2}$. The generating relation (Rodrigue's formula) for the Hermite polynomials is
\begin{equation}
\label{eq:Rodrigue}
(d/d\xi)^{n}\exp(-\xi^{2})=(-1)^{n}H_{n}(\xi)\exp(-\xi^{2}),
\end{equation}
and using this in Eq.~(\ref{eq:chin}) gives
\begin{equation}
\label{eq:chiHn}
\chi_{n}(x,t)\propto \frac{(t\!+\!\imath\tau)^{n}\gamma^{n}}{ \sqrt{ (t-\imath \tau)}}\exp [\frac{\frac{1}{2}\imath m x^{2}}{\hbar(t\!+\!\imath\tau)}]\,e^{- x^{2}\!/\gamma^{2}}\!H_{n}(\frac{x}{\gamma}).
\end{equation}
Separating the real and imaginary parts of the first exponent, the normalized sequence of solutions to SE is
\begin{equation}
\label{eq:chiHtheta}
\chi_{n} (x,t)=\frac{\pi^{-1/4}}{ \sqrt{ \gamma\, 2^{n}n!}}e^{\imath (\theta +(n-1)\beta)}\,e^{- x^{2}/2\gamma^{2}}H_{n}(\frac{x}{\gamma}),
\end{equation}
where $\theta= t x^{2}/2\tau \gamma^{2}$ and $e^{\imath\beta}=[(t-\imath \tau)/(t+\imath \tau)]^{1/2}$. The probability distributions of the first few are:
\begin{eqnarray}
\label{eq:chiProb}
|\chi (x,t)|^2 & = &\frac{1}{\gamma \sqrt{\pi}} \,\exp (- x^2/\gamma^2) \\
|\chi_{1} (x,t)|^2 & =  &\frac{2}{\gamma \sqrt{\pi}} \,\frac{x^2}{\gamma^2}\exp (- x^2/\gamma^2) \\
|\chi_{2} (x,t)|^2 & =  &\frac{1}{2\gamma \sqrt{\pi}} \,(1-2x^2/\gamma^2)^2 \exp (- x^2/\gamma^2). 
\end{eqnarray}
These do not change their shape as they evolve; only the scale of $x$ changes with time. It would serve little purpose to show a graph of these probability distributions because they have the same spatial form as do the energy eigenfunctions of the harmonic oscillator. The relation to the harmonic oscillator is made clearer in Section VII.

The fact that the Hermite-Gauss packets do not change shape is consistent with the asymptotic results in Section III because the Fourier transform of a Hermite-Gaussian is also a Hermite-Gaussian.
\begin{equation}
\label{eq:FT1}
\frac{1}{\sqrt{2\pi}}\int_{-\infty}^{\infty}\!e^{-z^{2}/2}H_{n}(z)e^{-\imath p z}dz= \imath^{-n}e^{-p^{2}/2}H_{n}(p).
\end{equation}

\section{Examples}

\begin{figure}
\includegraphics[]{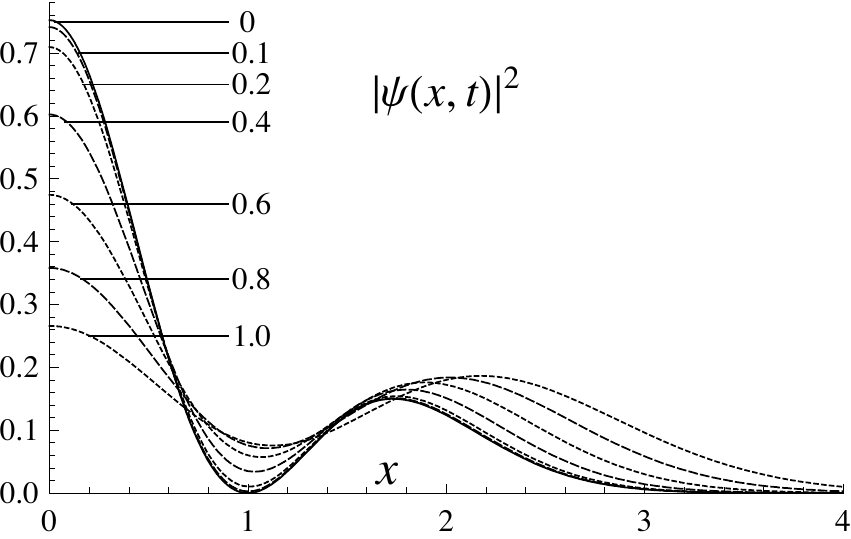}
\caption{\label{fig:Fig1} The probability distribution $|\psi (x,t)|^{2}$ for the wave packet in Eq.(\ref{eq:chi2}), whose initial probability distribution is shown by the solid curve. Only half of the packet is shown, because it remains symmetric. The probability is shown in the period before the asymptotic regime for the times $t=0, 0.1, 0.2, 0.4, 0.6, 0.8, 1.0$ in units of $\tau $. The distance $x$ is in units of $(\hbar\tau/m)^{1/2}$.}
\end{figure}

Our first example is a wave packet which is smooth (infinitely differentiable) everywhere and goes to zero exponentially at large distances, but does change shape as it evolves. We use the second derivative of Schr\~odinger's packet $\chi$ in Eq.(\ref{eq:chi}). Thus $\bar{\chi}_{2}(x,t)\propto\partial_{x}^{2}\chi(x,t)$, which also must satisfy SE. It has the form
\begin{equation}
\label{eq:chi2}
\bar{\chi}_{2}(x,t)=N\kappa^{3}(2\kappa^{2}x^{2}-1)e^{-\kappa^{2}x^{2}},
\end{equation}
where $\kappa^{2}=-\imath m/2\hbar (t-\imath \tau)$ and $N$ is the unimportant normalizing constant $(3\sqrt{\pi}/32)^{-\frac{1}{2}}$. Initially, the wave function is real. It is also symmetric about $x=0$ and will maintain that symmetry. As shown in Fig.~\ref{fig:Fig1} there is a hump centered on the origin with a lower hump on either side. We will see that as it evolves, the central hump diminishes and eventually disappears while the outer humps move further out and spread. The initial packet has $\Delta_{x}^{2}=7\hbar\tau/6m$ and $\Delta_{p}^{2}=5\hbar m/2\tau$, and hence $t_{p}=\tau/5$ and $t_{x}=7\tau/3$. We expect that the shape (of the probability distribution) will not change significantly over periods much less than $t_{p}$. Initially, no change can be seen in the graph until $t\approx t_{p}/3\approx 0.06\tau$. The maximum rate of change appears to be at about $t=0.6$ and $x=0$ when visible change starts for $\delta t \approx t_{p}/30$. 

\begin{figure}
\includegraphics[]{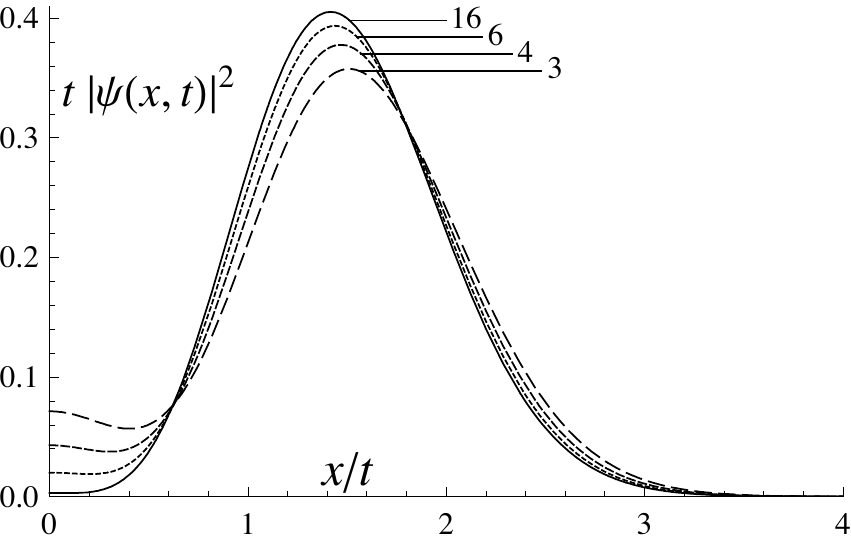}
\caption{\label{fig:Fig2} The probability distribution for the same wave packet as in Fig.~\ref{fig:Fig1} for some times $(t=3, 4, 6, 16)$ when the packet is approaching its asymptotic form. In the asymptotic region, $t|\psi (x,t)|^{2}$ is a function of $x/t$ only, as in Eq.(\ref{eq:chi2a}), so we use $x/t$ as the horizontal variable in this figure, which hides the substantial spreading of these packets. For all $t> 7t_{x}\approx 16 \tau$ the exact distribution is indistinguishable from the asymptotic form; so the graph for $t=16$ stands for all later times.}
\end{figure}

Taking the Fourier transform of $\bar{\chi}_{2}$ leads to the asymptotic form
\begin{equation}
\label{eq:chi2a}
\bar{\chi}_{2}(x,t)\approx N'\frac{x^2}{t^{5/2}}\exp\big(-\frac{m \tau  x^2}{2h t^2}\big),
\end{equation}
with $N'=2(m^5 \tau^5/9\pi\hbar^5)^{1/4}$, which we expect to be valid for $t\gg t_{x}$. No error in this approximation can be seen in Fig.~\ref{fig:Fig2} after $t\approx 7t_{x}$.

\begin{figure}
\includegraphics[]{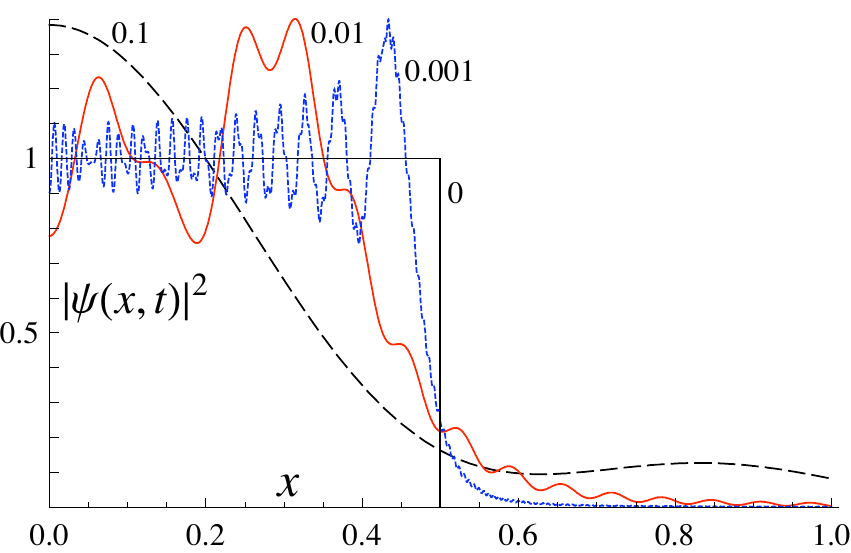}
\caption{\label{fig:Fig3} The probability distribution $|\psi (x,t)|^{2}$ for the initially square wave packet in Eq.(\ref{eq:rect}), for several values of the time $(t=0, 0.001, 0.01, 0.1)$ in the early period when the packet shows complicated behavior, but does not spread appreciably. The distance $x$ is in units of $a$, the width of the initial packet. The times are in units of $ma^{2}/\hbar$ and are indicated by the numbers against the graphs.}
\end{figure}

\begin{figure}
\includegraphics[]{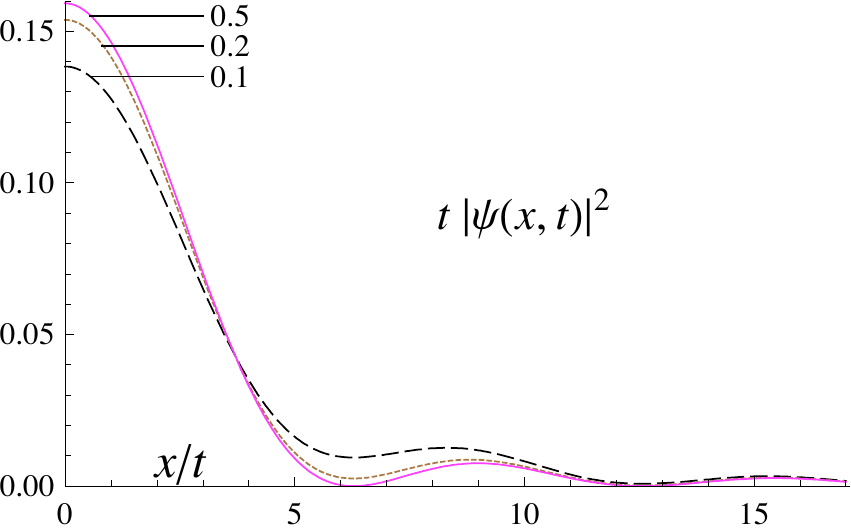}
\caption{\label{fig:Fig4} The probability distribution for the initially square packet for some times $(t=0.1, 0.2, 0.5)$ when the packet is approaching its asymptotic form. We show $t|\psi (x,t)|^{2}$ as a function of $x/t$. For all $t\geqslant 0.5$ the exact distribution is indistinguishable from the asymptotic form; so the graph for $t=0.5$ stands for all later times.}
\end{figure}

As a second example, consider the square wave packet, with 
\begin{equation}
\label{eq:rect}
 \psi(x,0)=  \begin{cases} 1/\sqrt{a},   &|x|<a/2 \\ 0,   &|x|\geq a/2.\end{cases}
\end{equation}
The momentum wave function is, from Eq.~(\ref{eq:mom}),
\begin{equation}
\label{eq:phiRect}
 \phi(p)=\sqrt{\frac{a}{2\pi \hbar}}\frac{\sin (a p/2\hbar)}{a p/2\hbar}
\end{equation}
and therefore, for $t\gg ma^2 /\hbar$, 
\begin{equation}
\label{eq:appRect}
 \psi(x,t)\approx \sqrt{\frac{a m}{2\pi \imath \hbar t}}\, \exp [\frac{\imath m x^2 }{ 2\hbar t}]\frac{\sin (a m x / 2\hbar t)}{a m x / 2\hbar t}.
\end{equation}
The exact evolution can be carried out, using Eq.~(\ref{eq:intProp}), in terms of error functions with complex argument, or Fresnel functions with real argument. Fig.~\ref{fig:Fig3} shows the exact probability distribution for some times well before it reaches the asymptotic realm. Fig.~\ref{fig:Fig4} shows the exact probability distribution for some times when the exact distribution is approaching its asymptotic form. 

In this case the initial value of $\Delta_{x}^{2}$ is $a^{2}/12$ so that $t_{x}=ma^{2}/6\hbar$. Fig.~\ref{fig:Fig4} shows that the asymptotic form is a close approximation for $t>3t_{x}$. The short-period theory in Section II does not apply because $\Delta_{p}=\infty$, and there are rapid changes during the early evolution. Discontinuous wave functions are unphysical; an infinite energy would be required to produce them. This does not mean that they cannot be useful as mathematically simple states that can be closely approximated by physical ones.

\newpage
\section{Further considerations}

The material above is simple enough and sufficiently complete to be presented in an elementary course in quantum mechanics, but some related topics may be helpful and may serve as extension material for students.

(a) For a more rigorous approach to the short-time behavior, write the exact form of Eq.~(\ref{eq:psipa}) as $\psi(x,t)=\exp(-\imath \bar{p}^{2}t/2m\hbar)\,\psi(x-\bar{p}t/m,0)+\delta \psi$, where
\begin{eqnarray}
\delta\psi(x,t)  =  \frac{1}{\sqrt{2\pi \hbar}}\exp(\frac{-\imath\bar{p}^{2}t}{2m\hbar})\times \hspace {3cm}\\
  \int_{-\infty}^{\infty}\!\!\!\!\exp [\frac{\imath}{\hbar} p(x-\frac{\bar{p}t}{m})]\,\big(\exp[\frac{-\imath(p-\bar{p})^{2}}{2m\hbar}]-1\big)\phi(p)\,dp. \nonumber
\end{eqnarray}
Now make use of the Cauchy-Schwarz inequality $|\int \!f(p)g(p)dp|^{2}\leqslant\int | f(p)|^{2}dp\int |g(p)|^{2}dp$, with $g(p)=(p-\bar{p})\phi(p)$ and $|f(p)|=|\sin[(p-\bar{p})^{2}t/4m\hbar]/(p-\bar{p})|$, and the integral $\int_{-\infty}^{\infty}\sin^{2}(z^{2})/z^{2}dz=\sqrt{\pi}$, to show that 
\begin{equation}
\label{ }
|\delta \psi|^{2}\leqslant\sqrt{t/\pi m \hbar^{3}}\Delta_{p}^{2}.
\end{equation}
This is a rigorous bound on the change in the wave function (apart from the shift $\bar{p}t/m$). For $|\delta \psi |$ to be small compared with $\psi$ with $|\psi |^{2}\sim 1/2\Delta_{x}$ (from the normalization of $\psi$), we require $t\Delta_{p}^{4}/\pi m \hbar^{3}\ll 1/4\Delta_{x}^{2}$ and hence $t\Delta_{p}^{2}/\pi m \hbar^{3}\ll 1/4\Delta_{x}^{2}\Delta_{p}^{2} \leqslant1/\hbar^{2}$ (from the uncertainty relation). That is, we require $t\ll \pi m \hbar/\Delta_{p}^{2}$, in agreement with Section II.

(b) For a more rigorous approach to the error in the asymptotic approximation, write the exact form of Eq.~(\ref{eq:large_t}) as
\begin{equation}
\label{ }
 \psi(x,t)= \sqrt{\frac{m}{ \imath t}}\, \exp \big[\frac{\imath m}{2\hbar t}(x^2-\bar{x}^{2})\big]\phi \big(\frac{m}{t}(x-\bar{x})\big)+\delta \psi\end{equation}
where
\begin{equation}
\label{ }
\delta\psi  =  \sqrt{\frac{m}{2\pi \imath \hbar t}}\, \exp [\frac{\imath m}{2\hbar t}(x^2-\bar{x}^{2})]\!\!\int_{-\infty}^{\infty}\!\!\!f(x')g(x')dx',
\end{equation}
with
\begin{eqnarray}
f(x') & \!\!= & \!\! \exp [\frac{\imath m}{\hbar t} (\bar{x}-x) x']\frac{\exp [-\frac{\imath m}{\hbar t}(x'-\bar{x})^{2}]-1}{x'-\bar{x}} \\
g(x') &  \!\!= &  \!\!(x'-\bar{x})\psi(x',0). 
\end{eqnarray}
Now the Cauchy-Schwarz inequality gives
\begin{equation}
\label{ }
|\delta \psi|^{2}\leqslant\sqrt{m^{3}/\pi \hbar^{3}t^{3}}\Delta_{x}^{2}.
\end{equation}
This is a rigorous bound to the error in the asymptotic approximation. For $|\delta \psi |$ to be small compared with $\psi$ we require $|\delta \psi |^{2}\ll1/2\Delta_{x}$, which leads to $t\gg (4/\pi)^{1/3}m\Delta_{x}^{2}$, in agreement with Section II.

If the wave packet has discontinuities, as in the square wave in Eq.~(\ref{eq:rect}), this derivation of the error is applicable only when the initial time is taken to be the instant when the discontinuities exist. At any other time, $\Delta_{x}$ does not exist (because $m^{2}d_{t}^{2}\langle \hat{X}^{2}\rangle=2\langle \hat{P}^{2}\rangle$, as in Section IV, and $\langle \hat{P}^{2}\rangle=\infty$.) The divergence of $\langle \hat{x}^{2}\rangle$ can easily be seen for the asymptotic wave function of the initially square packet from Eq.~(\ref{eq:appRect}).

(c) The relation of the Hermite-Gauss solutions for free particle motion to the energy eigenfunctions of the harmonic oscillator can be further elucidated as follows. Separating the real and imaginary parts of the exponent in $\chi$ in Eq.~(\ref{eq:chi}) gives
\begin{equation}
\label{eq:chiRIm}
\chi (x,t)=\frac{(m \tau/\pi \hbar)^{1/4}}{ \sqrt{ (t-\imath \tau)}}\exp (\frac{\imath t  x^{2}}{2\tau \gamma^{2}})\,\exp (-\frac{ x^{2}}{2\gamma^{2}}),
\end{equation}
with $\gamma = [\hbar (t^{2}+\tau^{2})/m\tau]^{1/2}$. Similarly to the derivation of Eq.~(\ref{eq:applyb}), one can show that
\begin{equation}
\label{eq:a}
e^{-\imath\theta}\,\hat b\,e^{\imath\theta}=\imath \sqrt{2\hbar m\tau}\,e^{\imath\beta}\,\hat a,
\end{equation}
where $\hat a=(\hat x/\gamma+\imath \gamma \hat p /\hbar)/\sqrt{2}$ and, as before, $\theta= t x^{2}/2\tau \gamma^{2}$ and $\beta=((t-\imath\tau)/(t+\imath\tau))^{1/2}$. Then $[\hat a,\hat a^{\dag}]=1$ and $\hat a$ is the type of operator familiar from the theory of the harmonic oscillator. Schr\~odinger's packet $\chi$, with $\hat b \,\chi =0$, corresponds to the ground state of $\hat a$, with $\hat a \exp (-x^{2}/2\gamma^{2})=0$. Then the sequence of states obtained by successive application of $\hat a^{\dag}$ to $ \exp (-x^{2}/2\gamma^{2})$ is the sequence of oscillator states $ \exp (-x^{2}/2\gamma^{2})H_{n}(x/\gamma)$ and corresponds to the sequence of solutions of the free SE obtained by successively applying $\hat b^{\dag}$ to $\chi$.

The Hermite-Gauss solutions can be readily generalized to have non-zero values of $\langle \hat x \rangle$ and $\langle \hat p \rangle$. Instead of $\hat b \,\chi =0$ leading to the solution for $\chi (x,t)$ in Eq.~(\ref{eq:chi}), start with $\hat b \,\chi=\beta \chi$, where $\beta = m\langle \hat x \rangle -\langle \hat p \rangle(t-\imath\tau)$, and then repeatedly apply $\hat b^{\dag}$.

It is not surprising that this Hermite-Gauss sequence of solutions exists for the free particle; a similar sequence can be found for any Hamiltonian that is at most quadratic in $\hat p$ and $\hat x$, even when the coefficients of the terms in the Hamiltonian vary with the time.\cite{And2}

(d) The spatial derivative of any solution of the free SE is also a solution. Taking successive derivatives of Schr\~odinger's packet $\chi$ in Eq.~(\ref{eq:chi}) gives a sequence of packets that also have the form of a Hermite polynomial multiplied by the original Gaussian and by a phase factor, but in this case the probability distributions do change shape as they evolve. 

From Rodrigue's Eq.~(\ref{eq:Rodrigue}), and defining $\bar{\chi}_{n}=\partial_{x}^{n}\chi$,
\begin{equation}
\label{eq:chinBar}
\bar{\chi}_{n}\,\,\propto\,\, \kappa^{n+1}e^{-\kappa^{2}x^{2}}H_{n}(\kappa x),
\end{equation}
where $\kappa=[m/2\imath\hbar(t-\imath\tau)]^{1/2}$. But the Fourier transform of this combination of a Gaussian and an exponential is different to that in Eq.~(\ref{eq:FT1}): \begin{equation}
\label{eq:FT2}
\frac{1}{\sqrt{2\pi}}\int_{-\infty}^{\infty}\!e^{-z^{2}}H_{n}(z)e^{-\imath p z}dz= \frac{(-\imath)^{n}}{\sqrt{2}}e^{-p^{2}/4}p^{n}.
\end{equation}
The momentum distribution of $\bar{\chi}_{n}$ is not a Hermite-Gaussian and therefore these packets change shape as they evolve. The first example in Section VII illustrates the case $n=2$.

\section{Conclusion}

To see the big picture of the free evolution of a wave packet, it helps to extend the evolution back to the distant past as well as forward. The centroid $\langle \hat{x}\rangle$ of the packet moves with constant speed $\langle \hat{p}\rangle/m$ and we know from Section IV that every packet will contract and then expand, so that $\Delta_{x}$ will take its minimum value $\Delta_{min}$ at some time $t_{0}$. Far from the minimum, for $|t-t_{0}|\gg t_{h}=m\Delta_{min}/\Delta_{p}$, $\Delta_{x}$ will vary linearly with time, from Eq.~(\ref{eq:delx}). We also know, from Section III, that for $|t-t_{0}|\gg t_{x}=2m\Delta_{min}^{2}/\hbar$ the shape of the wave packet will not change, but the scale will change linearly with time (consistent with the linear change in $\Delta_{x}$). Note that $t_{h}=(t_{x}t_{p})^{1/2}$, where $t_{p}=m\hbar/2\Delta_{p}^{2}$. Thus $t_{h}$ is the harmonic mean of $t_{x}$ and $t_{p}$ and $t_{p}\leqslant t_{h}\leqslant t_{x}$.

Therefore, changes in shape (other than in scale) can occur only in the waist region around $t_{0}$ and such changes cannot be rapid over times of the order of $t_{p}$, if $\Delta_{p}$ exists. But even in this waist region, there need not be changes in shape, as exemplified by the Hermite-Gauss packets.

\appendix*
\section{Comment on Mita's paper}
The only example considered in detail by Mita\cite{Mita} is the square wave packet, as in Eq.~(\ref{eq:rect}). We have shown that in this case, $|\psi(x,t)|$ evolves to approach the form $\sin(\beta x/t) /(\beta x/t)$ where $\beta=a m  / 2\hbar$, and this packet is not Gaussian. All that has been achieved in Mita's note is to impose a Gaussian fit to an arbitrary wave packet [by making a quadratic approximation to the logarithm in Mita's Eq.(14)]; if the packet is nothing like a Gaussian, the fit will be poor.

For Mita's parameters $m=\hbar=a=1$, the exact distribution differs from the approximate one [from Eq.~(\ref{eq:appRect})] by a negligible amount for all $t>0.5$. Mita's Fig. 3(c) shows only that, if you don't look too closely, $(\sin x/x)^2$ is not very different from a Gaussian shape; if the $x$-axis were extended further, the next maximum in the exact distribution would be revealed -- it does not decrease in relative size for later times.

\end{document}